\documentclass[sigconf, nonacm]{acmart}
\usepackage[subrefformat=parens]{subcaption}
\usepackage{tabularx}
\usepackage{layouts}

\usepackage[normalem]{ulem}

\newcommand\vldbdoi{XX.XX/XXX.XX}
\newcommand\vldbpages{XXX-XXX}
\newcommand\vldbvolume{14}
\newcommand\vldbissue{1}
\newcommand\vldbyear{2020}
\newcommand\vldbauthors{\authors}
\newcommand\vldbtitle{\shorttitle} 
\newcommand\vldbavailabilityurl{https://github.com/dbc-utokyoiis/DPFProto}
\newcommand\vldbpagestyle{plain} 

\begin{document}
\title{Data Path Fusion in GPU for Analytical Query Processing}

\author{Tsuyoshi Ozawa}
\orcid{0000-0002-4280-9559}
\affiliation{%
  \institution{The University of Tokyo}
  \streetaddress{Komaba 4-6-1}
  \city{Meguro-ku}
  \state{Tokyo}
  \country{Japan}
}
\email{ozawa@tkl.iis.u-tokyo.ac.jp}

\author{Kazuo Goda}
\orcid{0000-0003-0618-4157}
\affiliation{%
  \institution{The University of Tokyo}
  \streetaddress{Komaba 4-6-1}
  \city{Meguro-ku}
  \state{Tokyo}
  \country{Japan}
}
\email{kgoda@tkl.iis.u-tokyo.ac.jp}

\begin{abstract}
One major technical challenge for modern analytical database systems is how to leverage GPU to exploit their massive parallelism and high bandwidth.
Yet, existing GPU-driven database engines suffer from inefficiencies caused by frequent host-device interactions and fragmented execution across multiple GPU kernels, limiting their ability to fully utilize GPU's computational and IO capabilities.
This paper proposes Data Path Fusion (DPF), a novel GPU-driven data processing architecture that integrates a sequence of data path operations -- including IOs, decompression, and query operations -- into a single GPU kernel.
By fusing the data path, DPF reduces host-device communication overheads and enables more efficient utilization of GPU resources for analytical query workloads.
DPF seamlessly integrates GPU-friendly optimization techniques, including type-specific compression/decompression, variable-length attribute support, and state-of-the-art GPU-driven IO mechanism, to work in concert, enabling efficient end-to-end query execution directly on GPU.
Through extensive experimental evaluation using a prototyped DPF-based GPU-driven database engine (DPFProto) with analytical benchmark workloads, this paper demonstrates that DPF achieves speedups of 2.66 to 6.22 on TPC-H and 3.84 to 16.81 on SSB over the state-of-the-art approach in the representative configuration.
Our results show that DPF effectively unlocks the computational and IO potential of modern GPU, providing a promising direction for next-generation analytical database systems.
\end{abstract}

\maketitle

\pagestyle{\vldbpagestyle}
\begingroup\small\noindent\raggedright\textbf{PVLDB Reference Format:}\\
\vldbauthors. \vldbtitle. PVLDB, \vldbvolume(\vldbissue): \vldbpages, \vldbyear.\\
\href{https://doi.org/\vldbdoi}{doi:\vldbdoi}
\endgroup
\begingroup
\renewcommand\thefootnote{}\footnote{\noindent
This work is licensed under the Creative Commons BY-NC-ND 4.0 International License. Visit \url{https://creativecommons.org/licenses/by-nc-nd/4.0/} to view a copy of this license. For any use beyond those covered by this license, obtain permission by emailing \href{mailto:info@vldb.org}{info@vldb.org}. Copyright is held by the owner/author(s). Publication rights licensed to the VLDB Endowment. \\
\raggedright Proceedings of the VLDB Endowment, Vol. \vldbvolume, No. \vldbissue\ %
ISSN 2150-8097. \\
\href{https://doi.org/\vldbdoi}{doi:\vldbdoi} \\
}\addtocounter{footnote}{-1}\endgroup

\ifdefempty{\vldbavailabilityurl}{}{
\vspace{.3cm}
\begingroup\small\noindent\raggedright\textbf{PVLDB Artifact Availability:}\\
The source code, data, and/or other artifacts have been made available at \url{\vldbavailabilityurl}.
\endgroup
}

\section{Introduction}

Analytical query processing lies at the heart of modern data-driven enterprises,
where organizations increasingly demand real-time insights over massive datasets~\cite{DBLP:journals/cacm/ChaudhuriDN11}.
To meet these demands, analytical database systems must exploit every available hardware capability,
and Graphics Processing Units (GPUs) have emerged as a particularly compelling resource~\cite{DBLP:journals/micro/NickollsD10, DBLP:journals/micro/DallyKK21}.
Modern GPUs offer thousands of parallel execution units, memory bandwidths exceeding one terabyte per second, and dedicated support for high-throughput IOs~\cite{nvidia-a100-datasheet, nvidia-h100-datasheet, nvidia-gds-overview}.
These characteristics make GPUs a natural fit for the data-intensive, compute-regular workloads that dominate Online Analytical Processing (OLAP).

Despite this potential,
realizing GPU-native analytical query processing in practice has proven elusive.
Existing GPU-driven database engines typically decompose query execution into a pipeline of discrete GPU execution units,
each handling a specific stage of processing -- IO, decompression, filtering, aggregation, and so on.
While this modular design simplifies implementation,
it incurs substantial overheads that erode the very performance benefits that GPUs promise;
kernel launch latencies and synchronization barriers accumulate across stages, intermediate results must be materialized in GPU memory between kernels, and the CPU-centric control flow prevents GPU from sustaining its parallel pipeline.
Together, these inefficiencies mean that state-of-the-art GPU-driven analytical engines leave a significant fraction of GPU's compute and IO throughput on the table.
The recently proposed GOLAP architecture~\cite{GOLAP24} has pioneered the idea of placing GPU directly in the data path for OLAP workloads,
demonstrating that tight integration of IO and query processing on GPU opens a promising avenue for high-performance analytical systems.
Building on this foundation, this paper explores how far GPU-native execution can be pushed by unifying IO, decompression, and query operations into a single, seamlessly fused execution unit on GPU.

This paper answers that opportunity with \emph{Data Path Fusion} (DPF), a new GPU-driven data processing paradigm for analytical query processing.
The central insight of DPF is that all stages of the analytical data path — IO, decompression, and query operations — can and should be fused into a \emph{single GPU kernel}.
By unifying these stages within one kernel,
DPF eliminates inter-kernel synchronization barriers and intermediate result materialization,
keeping execution entirely on GPU from the moment data leaves storage to the moment query results are produced.
This fusion allows GPU to sustain its massively parallel throughput across the full data path without CPU intervention.

DPF further incorporates a suite of GPU-friendly optimization techniques,
each designed to work in concert with the kernel fusion architecture.
\emph{Type-specific compression/decompression} applies attribute-level, type-aware encoding within each database page,
reducing the volume of data that must traverse
the secondary storage and the GPU memory hierarchy and improving overall compression efficiency.
\emph{Variable-length attribute support} enables DPF to handle real-world analytical schemata containing strings and other variable-length fields,
broadening its applicability beyond fixed-length data types.
\emph{BaM-based IO mechanism}~\cite{BAM23} provides fine-grained, GPU-thread-driven storage access that closely matches the access patterns of fused kernel execution,
further elevating effective IO bandwidth.
Together, these techniques allow DPF to fully realize the computational and IO potential of modern GPUs for analytical query workloads.

The core contributions of this paper are as follows.
\begin{enumerate}
    \item This paper proposes a new GPU-driven data processing paradigm called \emph{Data Path Fusion} (DPF) for analytical query processing.
    The \emph{kernel fusion} architecture allows executing a sequence of data path operations -- IO, decompression, and database operations -- within a single GPU kernel,
    thereby reducing host-device communications
    while improving the utilization of GPU's computational capacity and IO bandwidth for analytical query processing.
    \item This paper presents GPU-friendly optimization techniques --
    including type-specific compression/decompression, variable-length attribute support, and BaM-based IO --
    which work in concert with the kernel fusion architecture,
    enabling efficient end-to-end query execution directly on GPU.
    \item This paper reports intensive and extensive experiments
    demonstrating that DPF allows a GPU-driven database engine to perform significantly faster for analytical query workloads compared to existing approaches.
\end{enumerate}

The rest of this paper is organized as Sections 2 to 7, covering background and highlights, design, implementation, evaluation, related work, and conclusion.

\begin{table*}[t!]
    \centering
    \caption{Major work on GPU-driven storage-class database engines.
    \textit{DPF distinguishes itself by incorporating the kernel fusion architecture with type-specific compression, variable-length attribute support and BaM-based IO. In addition, our work provides transparent loading cost evaluation.}
    }
    \label{tab:majorwork}
    \begin{tabularx}{\linewidth}{l|X|X|X|X}
      \toprule
       &
      \multicolumn{1}{c|}{\textbf{Data Path Fusion (Ours)}} &
      \multicolumn{1}{c|}{Proteus + GDS~\cite{DBLP:conf/damon/NicholsonCBA25}} &
      \multicolumn{1}{c|}{GOLAP~\cite{GOLAP24}} &
      \multicolumn{1}{c}{HippogriffDB~\cite{DBLP:journals/pvldb/LiTLPS16}} \\
      \midrule
      Kernel control &
      \textbf{Fusion of IO, decompression and query operations} &
      Fusion of query operations &
      Fusion of query operations\textsuperscript{\dag1} &
      Fusion of query operations \\
      Compression scheme &
      \textbf{Type-specific scheme decision} &
      Multiple schemes evaluated\textsuperscript{\dag2} &
      Scheme optimization at loading\textsuperscript{\dag3} &
      Scheme optimization at loading and query processing\textsuperscript{\dag4} \\
      IO mechanism &
      \textbf{GPU-initiated (BaM)} &
      CPU-initiated (GDS) &
      CPU-initiated (GDS) &
      CPU-initiated (Hippogriff) \\
      Pruning &
      \textbf{GPU-side pruning} &
      Not reported &
      GPU-side pruning &
      Not reported \\
      Variable-length attribute &
      \textbf{Explicitly proposed and evaluated} &
      Explicitly proposed (not evaluated) &
      Not proposed &
      Evaluated in two test queries\textsuperscript{\dag5} \\
      Loading cost &
      \textbf{Evaluated} &
      Not reported &
      Not reported &
      Not reported \\
      \bottomrule
    \end{tabularx}
    \vspace{1mm}
    {\footnotesize
    \raggedright\textsuperscript{\dag1}
    The original paper describes that multiple different columns are processed in a single kernel, implying the deployment of operator fusion.
    \raggedright\textsuperscript{\dag2}
    Multiple compression schemes are intensively evaluated, but the decision method at query time is not discussed.
    \raggedright\textsuperscript{\dag3}
    Multiple compression schemes are evaluated and one of them is chosen at loading time.
    \raggedright\textsuperscript{\dag4}
    Multiple compressed versions are generated at loading time and one of them is utilized for query processing.
    \raggedright\textsuperscript{\dag5}
    A design for supporting variable-length attributes is not disclosed, but two test queries containing variable-length strings are found in their experiments\cite{DBLP:journals/pvldb/LiTLPS16}.
    \par
    }
\end{table*}

\section{Background and Highlights}
\label{sec:background}

\subsection{GPU and Analytical Database Engines}

The idea of harnessing GPU parallelism for analytical data processing has inspired a rich and sustained body of research in the database community,
with contributions spanning more than two decades.

The foundational efforts established that GPU could meaningfully accelerate the core operators of relational query processing.
Govindaraju et al. took an early and influential step by implementing conjunctive selections, aggregations,
and semi-linear queries on GPU's graphics pipeline~\cite{DBLP:conf/sigmod/GovindarajuLWLM04},
and subsequently demonstrated the potential of GPU-based external sorting through GPUTeraSort on the Sort Benchmark~\cite{DBLP:conf/sigmod/GovindarajuGKM06}.
The same group examined GPU-accelerated relational joins in depth~\cite{DBLP:conf/sigmod/HeYFLGLS08} and culminated this line of work in GDB, a GPU query engine supporting selections, aggregations, joins, and sorting, evaluated end-to-end on TPC-H~\cite{DBLP:journals/tods/HeLYFGLS09}.
On the data representation front, Fang et al. ported lightweight compression schemes -- including null suppression, dictionary coding, RLE, and bit packing -- to GPU~\cite{DBLP:journals/pvldb/FangHL10},
and Satish et al. analyzed sorting performance under bandwidth-oblivious SIMD formulations across CPU and GPU~\cite{DBLP:conf/sigmod/SatishKCNLKD10}.
These pioneering works collectively established the viability of GPU-based query processing and laid the operator-level foundation upon which subsequent system builders would construct.

Motivated by these operator-level successes, the community turned to constructing full-fledged GPU-accelerated analytical database systems~\cite{DBLP:journals/pvldb/PirkMZM16, DBLP:journals/pvldb/ChrysogelosKAA19,DBLP:conf/sigmod/FunkeBNMT18,DBLP:conf/sigmod/ShanbhagMY20,DBLP:journals/pvldb/YogatamaGY22,DBLP:journals/pvldb/HongKLMBH24}.
An important insight emerged early in this phase:
Yuan et al. quantified that while GPU excels at compute-bound queries, PCIe transfer overhead frequently dominates end-to-end latency~\cite{DBLP:journals/pvldb/YuanL013},
shifting attention from raw compute throughput to the host-device bandwidth bottleneck.
Researchers responded with a variety of creative approaches.
Ocelot introduced hardware-oblivious parallelization for in-memory column stores~\cite{DBLP:journals/pvldb/HeimelSPMM13},
HyPE developed a self-learning cost model to route query operators adaptively between CPU and GPU~\cite{DBLP:journals/pvldb/Bress13},
and MultiQx-GPU managed GPU memory contention across concurrently executing queries~\cite{DBLP:journals/pvldb/WangZYMLD014}.
Together, these works reveal the community's growing awareness that GPU-accelerated analytics is as much a systems integration challenge as a parallel computing one.

As working sets grew beyond GPU memory, researchers began to confront the full storage-to-GPU data path.
HippogriffDB~\cite{DBLP:journals/pvldb/LiTLPS16} was an early and important step in this direction, transferring data directly from NVMe SSDs to GPU via peer-to-peer pathways that bypass host memory,
while decompressing data on GPU at query time.
HippogriffDB also introduced \emph{operator fusion}, which merges multiple relational operators into a single GPU kernel to eliminate intermediate results and improve kernel efficiency.
These insights underscored that effective IO management is as critical as compute efficiency in GPU-driven analytics.

In parallel with these architectural explorations, advances in GPU hardware -- larger core counts, expanded shared memory, and richer warp-level primitives -- continued to reshape operator design.
Notable contributions include GPU-accelerated string matching for SQL queries~\cite{DBLP:journals/vldb/SitaridiR16},
exploitation of high-speed interconnects such as NVLink and RDMA for multi-GPU and distributed joins~\cite{DBLP:conf/sigmod/LutzBZRM20,DBLP:conf/sigmod/LutzBZRM22,DBLP:conf/sigmod/PaulLHL21,DBLP:journals/pacmmod/ThostrupDBLB23},
and tile-based lightweight integer compression designed specifically for GPU execution characteristics~\cite{DBLP:conf/sigmod/ShanbhagYYM22}.

Building on all of these advances, GOLAP~\cite{GOLAP24} recently proposed a GPU-in-Data-Path (GiDP) architecture that places GPU directly on the critical path between storage and query results.
By combining SSD-to-GPU transfer via GPUDirect Storage with GPU-side IO pruning based on block-level summaries and on-the-fly decompression,
GOLAP demonstrated that a tightly integrated GiDP is a practical and promising direction for high-performance OLAP.
Our work, \emph{Data Path Fusion} (DPF), is inspired by and builds upon this vision, and asks how much further this integration can be pushed through \emph{kernel fusion} of the entire data path -- unifying not only database operators but also IO and decompression into a single GPU kernel, extending the philosophy of operator fusion to encompass the full journey from storage access to query result production.

\subsection{Highlights on Our Work}

This paper proposes \emph{Data Path Fusion} (DPF),
a GPU-driven data processing paradigm that pushes the integration of analytical query execution on GPU to its logical conclusion.
Table~\ref{tab:majorwork} summarizes the key characteristics of DPF alongside recent work on GPU-driven storage-class database engines.
We highlight the distinguishing aspects of DPF below.

\noindent
\textbf{Kernel fusion.}
DPF fuses an entire sequence of data path operations -- IO, decompression, and database operations -- into a single GPU kernel.
By keeping the full data path within one kernel,
DPF minimizes host-device interactions that otherwise arise from orchestrating multiple discrete kernels,
thereby reducing synchronization overheads and intermediate materialization and allowing GPU to fully exploit its computational capacity and memory bandwidth.

\noindent
\textbf{Type-specific compression/decompression.}
DPF applies compression schemes tailored to the data type of each column, rather than treating all columns uniformly.
This type-aware, attribute-level encoding achieves higher compression ratios, reducing the volume of data transferred from secondary storage to GPU and traversing the GPU memory hierarchy during query execution.

\noindent
\textbf{BaM-based IO.}
DPF adopts BaM~\cite{BAM23} as its GPU-driven IO mechanism,
which allows IO requests to be issued and completed directly within the GPU kernel.
This design is inherently compatible with the kernel fusion architecture:
the fused kernel can initiate fine-grained storage accesses and immediately process the returned data without yielding control to the host, sustaining high IO throughput throughout query execution.

\noindent
\textbf{Variable-length attribute support.}
DPF supports variable-length attributes such as strings (\texttt{varchar}),
enabling it to handle real-world analytical schemata that go beyond fixed-length numeric types.
This broadens the applicability of GPU-driven storage-class query processing and allows DPF to demonstrate its effectiveness under more realistic database workloads.

\noindent
\textbf{Loading cost evaluation.}
Beyond query execution performance,
this paper also evaluates the cost of organizing data into the DPF storage format.
Loading is a practical concern for any database engine that employs specialized on-storage layouts or compression schemes,
and our experiments quantify this overhead to provide a complete picture of DPF's end-to-end practicality.

DPF distinguishes itself by bringing all of these capabilities together in a single, coherent architecture:
kernel fusion of data path with type-specific compression/decompression and BaM-based IO, and variable-length attribute support.
In addition, this paper provides transparent loading cost evaluation.
The following sections detail the design, implementation, and experimental evaluation.

\begin{figure*}[t!]
  \centering
  \includegraphics[width=\linewidth]{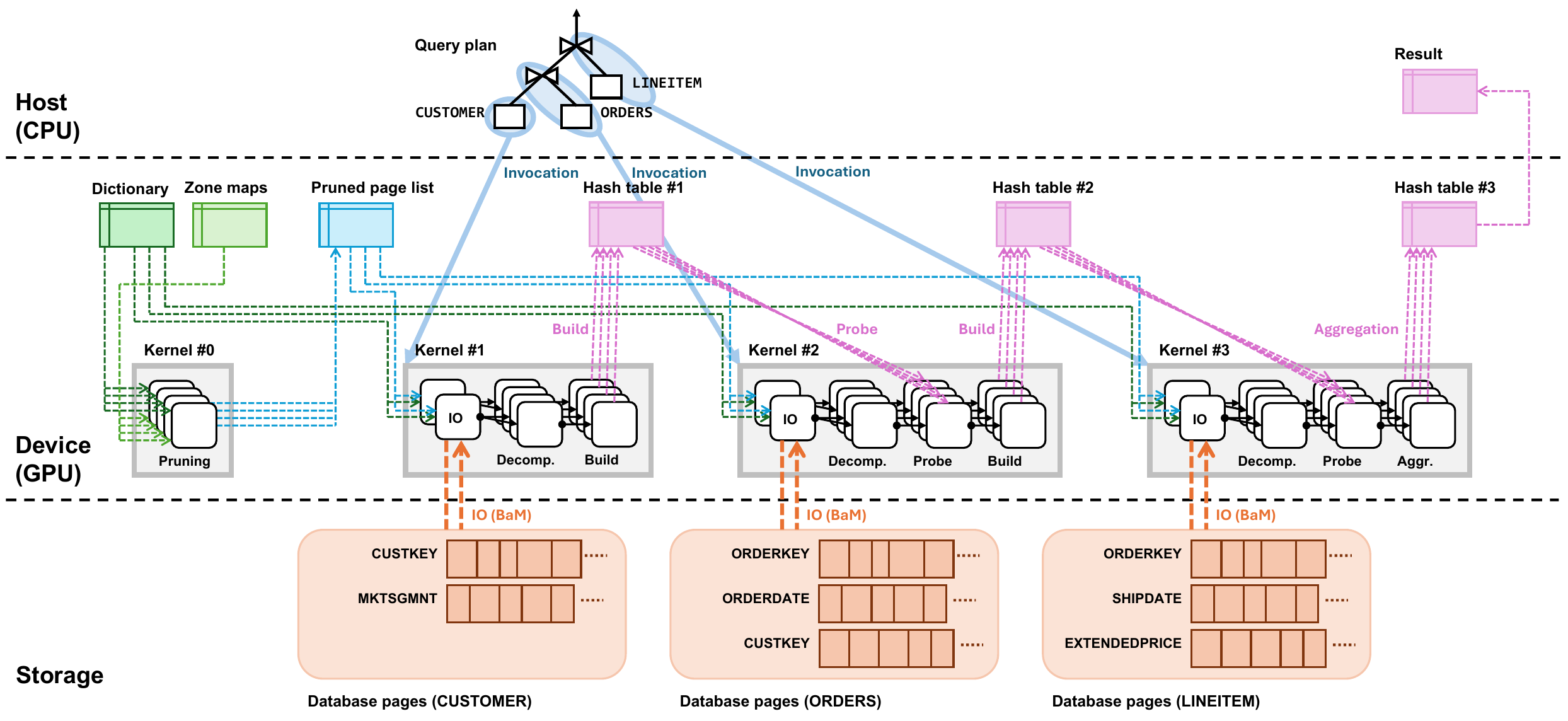}
  \caption{
  Overall architecture of Data Path Fusion (DPF).
  \textit{
  The diagram presents an example of a three-table join query over \textsf{CUSTOMER}, \textsf{ORDERS}, and \textsf{LINEITEM}.
  Each fused GPU kernel integrates BaM-based IO, decompression, and database operations within a single kernel,
  reducing host-device interactions.
  }
  }
  \label{fig:systemoverview}
\end{figure*}

\section{Data Path Fusion in GPU}
\label{sec:design}

This section presents the design of Data Path Fusion (DPF).
We first describe the overall architecture using a concrete query example, then detail each of the key components:
kernel fusion, BaM-based IO, and type-specific compression and decompression supporting variable-length attributes.

\subsection{Overall Design}
\label{sec:design:overall}

DPF is a GPU-driven data processing paradigm in which IO, decompression, and subsequent database operations are fused into a single GPU kernel and executed as one unit.
By integrating IOs with in-kernel decompression and database operations,
DPF avoids materializing intermediate results in global memory and eliminates the host-device synchronization overhead that arises when these stages are executed as separate kernels.

Figure~\ref{fig:systemoverview} illustrates the overall architecture of DPF using a three-table join query over \textsf{CUSTOMER}, \textsf{ORDERS}, and \textsf{LINEITEM} as a running example.
In receiving a query, the host computer generates a query plan and processes the query on GPU.
The core part of GPU-side query execution proceeds through three GPU kernel invocations (\textbf{Kernels~\#1 to \#3}), each being associated with a separate query operation pipeline.

\noindent
\textbf{Kernel~\#1.}
The first fused kernel works for the first query operation pipeline of scanning the \textsf{CUSTOMER} table and building a hash table.
Using the dictionary and the pruned page list,
the kernel identifies the storage locations of compressed pages for the relevant
columns (e.g., \textsf{CUSTKEY} and \textsf{MKTSEGMENT}) and issues BaM IO requests to read them.
Note that this paper assumes columnar storage layouts~\cite{DBLP:conf/vldb/StonebrakerABCCFLLMOORTZ05,DBLP:conf/sigmod/CopelandK85}, typical for modern analytical database engines, where columns are stored separately in storage.
Multiple columns can be read in parallel within the same kernel.
After ensuring IO completion,
the kernel decompresses the retrieved pages to retrieve decompressed column values, and builds a hash table (Hash table~\#1) that will be probed by the next kernel.

\noindent
\textbf{Kernel~\#2.}
The second kernel works for the second pipeline of scanning \textsf{ORDERS}, probing the hash table built in the previous kernel and building another hash table.
Following the same IO--decompress--operate pattern,
the kernel reads and decompresses the relevant columns (e.g., \textsf{ORDERKEY}, \textsf{ORDERDATE}, \textsf{CUSTKEY}) using the pruned page list.
The decompressed data is used to probe Hash table~\#1, and matching tuples are inserted into a new hash table (Hash table~\#2) for the next stage.

\noindent
\textbf{Kernel~\#3.}
The third kernel works for the last pipeline of scanning \textsf{LINEITEM}, probing the hash table built in the previous kernel and generating an aggregation result.
This kernel reads and decompresses columns (e.g., \textsf{ORDERKEY}, \textsf{SHIPDATE}, \textsf{EXTENDEDPRICE}), probing Hash table~\#2, and performing a hash-based aggregation to compute the final query result.
The aggregated result is then converted into a record-oriented format and returned to the host.

In this manner, each fused kernel encapsulates the full data path -- from storage IO to the production of its output -- within a single GPU kernel.
The host's role is limited to launching kernels and does not intervene during their execution, allowing GPU to sustain high throughput across the entire data path.

\subsubsection{GPU-side Pruning}

DPF employs a compile-time pruning technique similar to major GPU-oriented work~\cite{DBLP:journals/pvldb/YogatamaGY22, DBLP:journals/pvldb/YogatamaGY24}.
As Figure~\ref{fig:systemoverview} illustrates,
the fused kernel in charge of each query pipeline assumes that the pruned page list is made ready by a precedent stage (\textbf{Kernel~\#0}).
When a query arrives and the host computer generates a query plan,
DPF first performs pruning on GPU using dictionary and zone maps that are loaded into the GPU memory from storage.
Following the popular implementation practice~\cite{DBLP:conf/vldb/Moerkotte98,DBLP:journals/pvldb/LambFVTVDB12,DBLP:journals/pvldb/RamanABCKKLLLLMMPSSSSZ13,netezza-architecture},
this paper assumes that the dictionary tells pages storing data of each column, and the zone map tells key ranges of the column stored in each page;
key ranges includes maximum and minimum values of the column stored in the page, and those of other columns that can be joined with it~\cite{GOLAP24,10.1145/3722212.3724447,DBLP:journals/pvldb/RamanABCKKLLLLMMPSSSSZ13}.
Evaluating predicates appearing in the query pipeline with the dictionary and the zone map,
this stage produces a pruned list of pages that the query processing will access for each column,
eliminating pages that cannot satisfy the query predicate from the list, and
avoiding unnecessary storage IOs in later fused kernels.

\subsection{Kernel Fusion}
\label{sec:design:kernelfusion}

\begin{figure}[t!]  
  \centering
  \includegraphics[width=\linewidth]{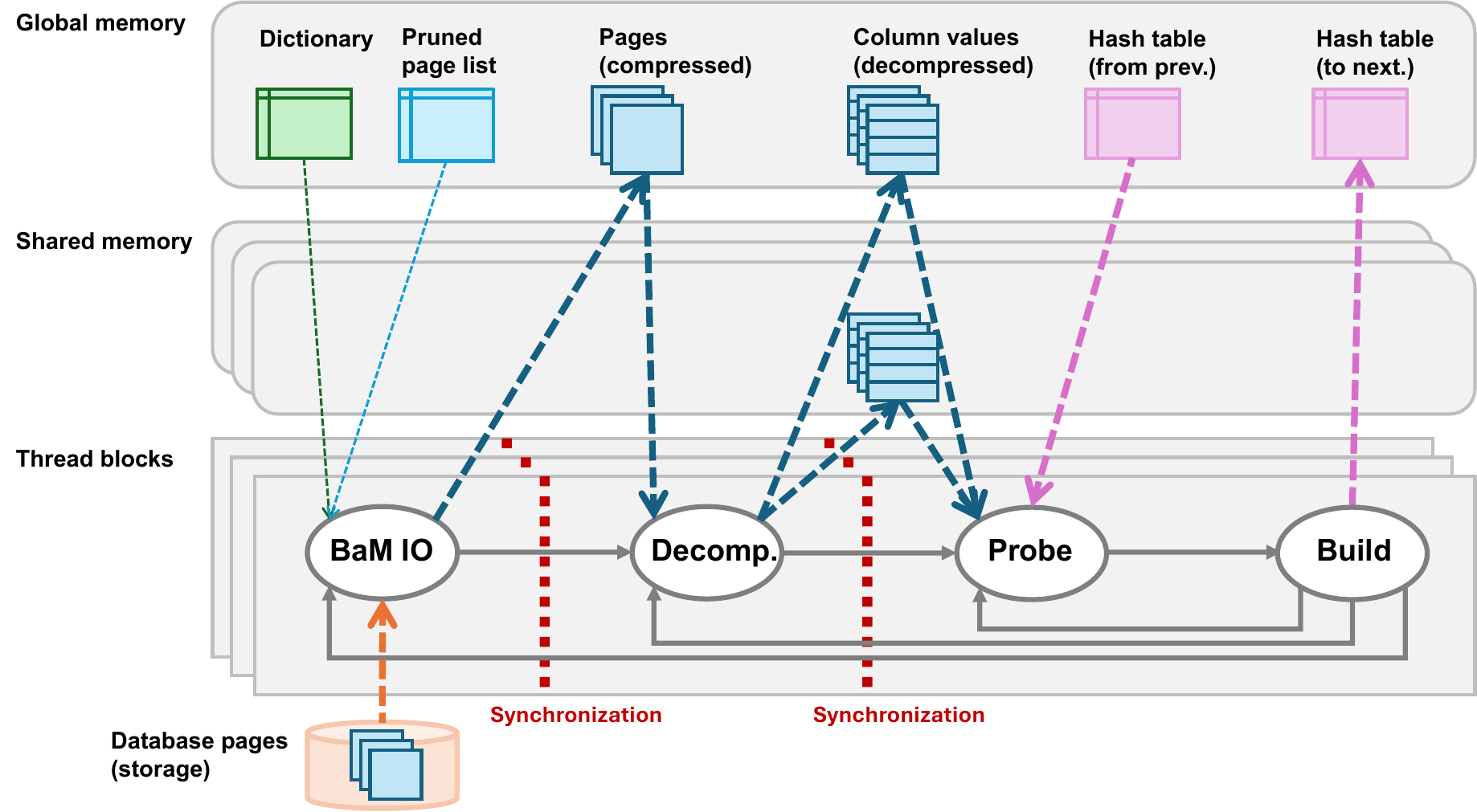}
  \caption{
  Internal structure of a fused GPU kernel.
  \textit{
  The diagram illustrates \textbf{Kernel~\#2} as an example.
  }
  }
  \label{fig:fusedkernel}
\end{figure}

The fundamental unit of execution in DPF is a fused kernel that combines IO, decompression, and one or more database operations into a single GPU kernel.
Figure~\ref{fig:fusedkernel} shows the internal structure of a fused kernel.

Within a fused kernel, thread blocks proceed through three stages separated by synchronization barriers.
In the \emph{IO stage},
threads issue BaM-based IO requests to read compressed pages from storage into global memory.
A synchronization barrier ensures that all IO operations have completed and their results are visible before proceeding.
In the \emph{decompression stage}, threads decompress the retrieved pages.
Decompressed column values are placed into shared memory (whenever the data volume fits within the available shared memory capacity) or global memory.
A second synchronization barrier ensures that all decompressed values are visible to every thread in the block.
In the \emph{operation stage}, threads execute the database operation -- such as hash table probing and building -- to produce its output in global memory.

The fused kernel has three iteration loops to feed IO and computational work to GPU without any interactions with the host computer.
The most-inner loop (i.e., within \emph{operation stage}) iterates the database operations until all the decompressed column values are processed.
The next inner loop (i.e., spanning \emph{decompression stage} and \emph{operation stage}) continues until all the read pages are processed.
The most-outer loop (i.e., spanning \emph{IO stage} to \emph{operation stage}) continues until all the necessary pages are processed.

\subsubsection{Thread organization and parallelism}

The degree of parallelism within a fused kernel is governed by the number of thread blocks and the number of threads per block.
In a typical configuration, DPF launches one thread block per Streaming Multiprocessor (SM) -- for example, 108 blocks on NVIDIA A100 -- with 128 threads (4 warps) per block, yielding 13,824 concurrently active threads.

Each stage of the fused kernel assigns thread roles as follows.
In the \emph{IO stage}, a designated subset of warps (e.g., warp~0 in each block) issues BaM IO requests and polls for completions,
while the remaining warps wait at the synchronization barrier.
Because each IO warp can occupy a dedicated NVMe queue pair, the number of IO warps directly controls the degree of storage-level parallelism.
In the \emph{decompression stage}, all threads participate:
each thread decompresses a small number of values -- four values per thread for GPU-FOR, one value per thread for FSST (later described in Section~\ref{sec:design:compression}).
The type-specific, value-level decompression granularity of DPF is what makes this fine-grained parallelism possible.
In the \emph{operation stage}, each thread processes one decompressed record (e.g., a hash table probe followed by an insertion), iterating until all records in shared memory have been consumed.
Both the decompression and operation stages thus utilize the full thread count of each block across all SMs.

DPF can adjust this configuration to match workload characteristics.
For queries that access many columns per table or involve compute-intensive operations, DPF increases the thread block size -- for example, to 1,024 threads (32 warps) per block,
yielding 110,592 active threads on A100.
In such configurations, a larger number of warps can be assigned to IO (e.g., 4 warps for 4 columns, enabling 432-way IO parallelism across all SMs), while the remaining warps handle decompression output that may spill to global memory due to the increased data volume.
The subsequent operation stage then benefits from the full 1,024 threads per block for compute-heavy tasks such as hash table probing and aggregation.
This flexibility -- trading shared memory residency for higher thread-level parallelism when downstream operations are compute-bound -- is a natural consequence of the fused kernel architecture,
which retains control over all stages within a single kernel.

\subsubsection{Grouping granularity}

One key design decision in DPF is the granularity at which columns are grouped into a single fused kernel.
DPF allows grouping one or more columns of the same table into one kernel rather than assigning a separate kernel to each column,
offering a clear technical benefit.
When multiple related columns are read and decompressed together,
their decompressed values can co-reside in shared memory,
enabling subsequent database operations to access all required columns without reading from or writing to the (relatively slower) global memory.

\subsection{BaM-based IO}
\label{sec:design:bam}

DPF adopts \emph{Big accelerator Memory} (BaM)~\cite{BAM23} as its GPU-initiated IO mechanism.
BaM enables GPU threads to issue IO requests directly from a running kernel, without returning control to the host CPU.
DPF uses BaM's internal asynchronous APIs to achieve fine-grained control over IO operations within the fused kernel.

At initialization time, BaM creates multiple NVMe queue pairs (submission and completion queues) for each storage device.
GPU threads access these queue pairs to submit IO requests and poll for completions.
This design offers flexibility in tuning IO concurrency.
In IO-intensive workloads, each warp can be assigned a dedicated queue pair to maximize IO throughput.
In compute-heavy workloads, the number of active queue pairs can be reduced to free GPU resources for computation.
The queue pair assignment is determined at kernel launch time and can be adjusted per kernel based on the characteristics of each stage of query execution.
This GPU-initiated IO model is inherently compatible with kernel fusion;
the fused kernel can seamlessly transition from the IO stage to decompression and database operations without host intervention%
\footnote{In contrast, CPU-initiated IO mechanisms such as GPUDirect Storage require the host to orchestrate each transfer, necessitating separate kernel invocations and host-device synchronization between the IO and processing stages.}.

\subsection{Type-specific Compression and Decompression}
\label{sec:design:compression}

DPF employs type-specific compression that leverages the data type of each column as defined by the database schema.
By selecting a compression scheme tailored to the type and value distribution of each column,
DPF achieves higher compression ratios than type-agnostic strategies,
reducing both the volume of data transferred from storage and the memory footprint within GPU.
A crucial design goal is to enable value-level decompression at fine granularity,
so that individual GPU threads can decompress small units of data,
aligning well with the fused kernel architecture.

\subsubsection{Fixed-length Attributes}
\label{sec:design:compression:fix}

\begin{figure}[t!]  
  \centering
  \includegraphics[width=\linewidth]{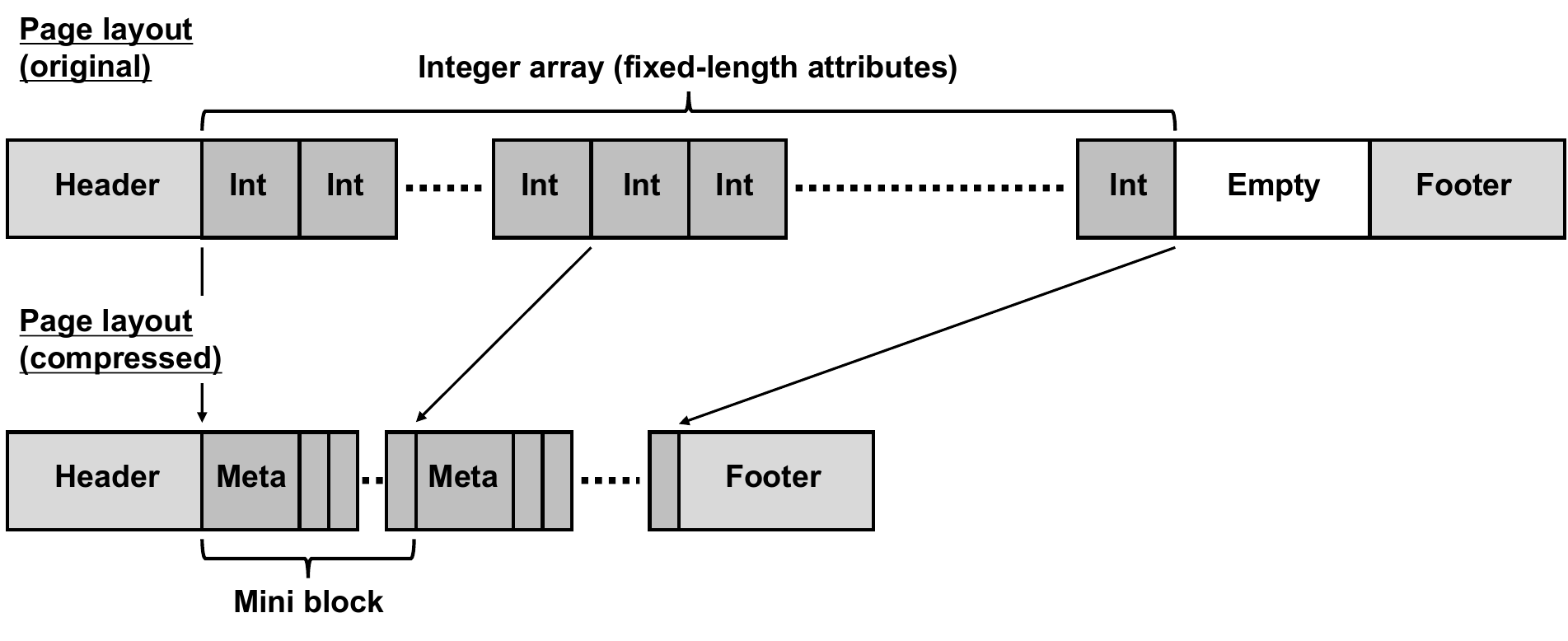}
  \caption{
  Page layout for fixed-length attributes.
  }
  \label{fig:fixedlength}
\end{figure}

Figure~\ref{fig:fixedlength} illustrates the page layout for fixed-length attributes in both original and compressed forms.
In the original layout (known as decomposed storage array~\cite{DBLP:conf/vldb/StonebrakerABCCFLLMOORTZ05,DBLP:conf/sigmod/CopelandK85}), a page consists of a header followed by a contiguous array of fixed-length column values and a footer.
In the compressed layout, the value array is partitioned into \emph{mini blocks}~\cite{DBLP:conf/sigmod/ShanbhagYYM22},
each preceded by metadata that records the information needed for decompression: a Frame-of-Reference (FOR) base value, the bit width of compressed values within the mini block, and the byte offset of the mini block within the page.

DPF selects the compression algorithm based on the column type.
32-bit integers are compressed using GPU-FOR~\cite{DBLP:conf/sigmod/ShanbhagYYM22};
64-bit integers use an extended 64-bit variant of GPU-FOR;
short fixed-length strings (\texttt{CHAR($n$)} where $n \le 2$) are reinterpreted as integers and compressed with GPU-FOR;
and longer fixed-length strings (\texttt{CHAR($n$)} where $n > 2$) are compressed as variable-length strings.

\subsubsection{Variable-length Attribute}
\label{sec:design:compression:variable}

Variable-length strings pose a challenge not present with fixed-length data:
because string lengths vary, the page containing a given row identifier (RID) cannot be determined from the RID alone by simple arithmetic.
DPF addresses this with two mechanisms:
an augmented page layout that embeds RIDs alongside string data, and an auxiliary RID index that enables efficient page lookup.

\begin{figure}[t!]  
  \centering
  \begin{minipage}[b]{1.0\linewidth}
      \includegraphics[width=\linewidth]{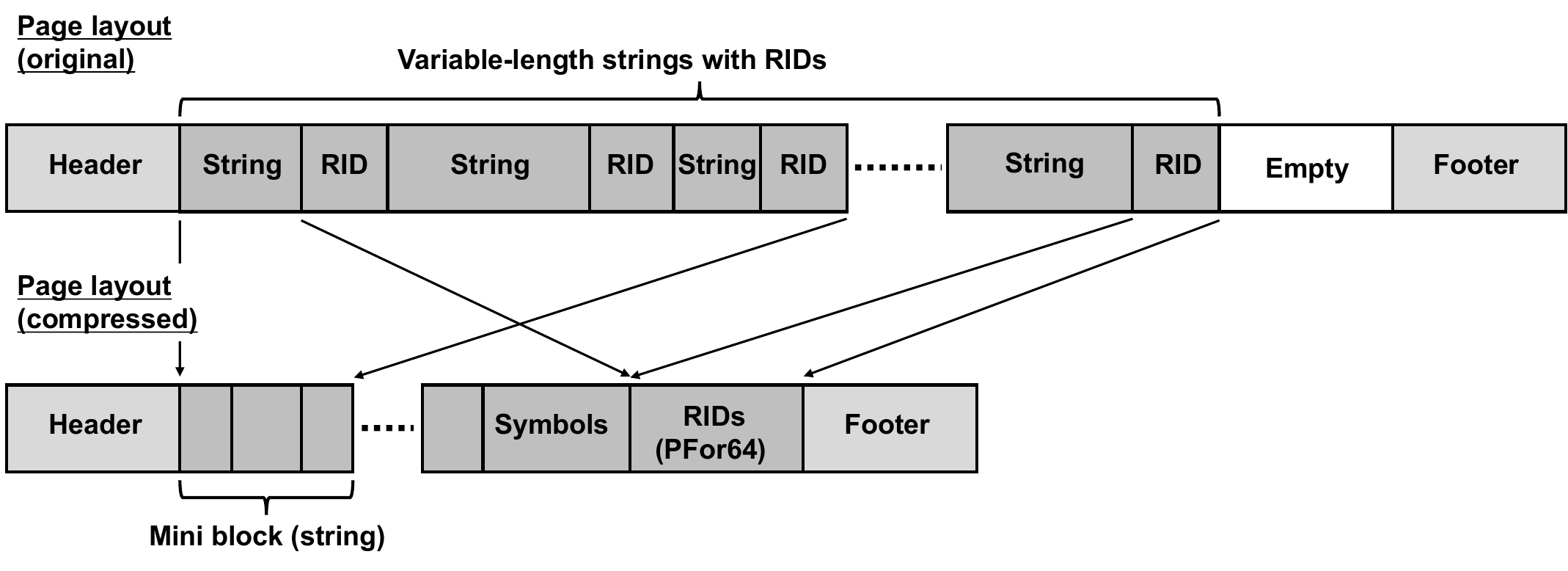}
      \subcaption{Page layout.}
      \label{fig:varcharpagelayout}
  \end{minipage}
  \begin{minipage}[b]{1.0\linewidth}
      \includegraphics[width=\linewidth]{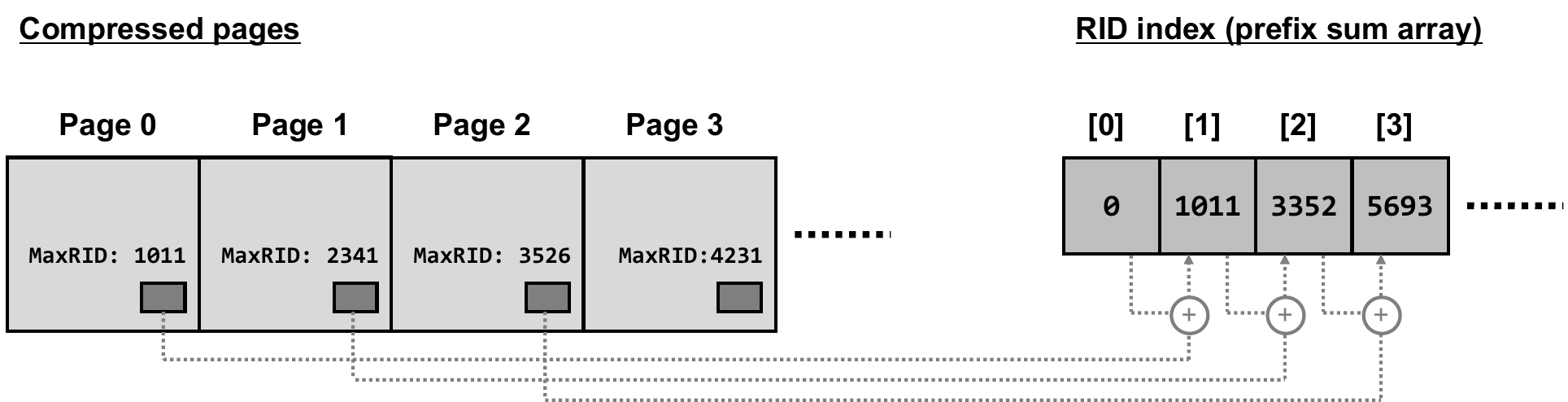}
      \subcaption{Row identifier (RID) index.}
      \label{fig:varcharrid}
  \end{minipage}
  \caption{
  Variable-length string storage and compression.
  }
  \label{fig:variablelength}
\end{figure}

Figure~\ref{fig:variablelength}~(a) shows the page layout for variable-length strings.
In the original layout, each page stores an interleaved sequence of string values and their associated RIDs.
In the compressed layout, string values are compressed using FSST~\cite{DBLP:journals/pvldb/Boncz0L20}.
Because compressed strings may span variable amounts of space,
the page may contain multiple string mini blocks,
each bounded by a configurable size limit%
\footnote{9 KB was set in our implementation because it allows a mini-block to fit in shared memory without sacrificing parallelism. Suppose 128 threads per block on A100; the per-SM limit of 2048 threads admits 16 blocks per SM, consuming 144 KB of shared memory -- within the 164 KB space.}.
A single FSST symbol table is stored per page and shared across all string mini blocks within that page.
The embedded RIDs are compressed separately using the 64-bit GPU-FOR scheme.

Figure~\ref{fig:variablelength}~(b) illustrates the RID index structure.
At data loading time, DPF constructs a \emph{RID index} for each variable-length column, which is a prefix-sum array over the number of records stored in each page.
Entry $i$ of the array stores the cumulative record count up to and including page~$i$;
the page containing a given RID can be located via binary search.
Although designed primarily for variable-length columns,
the RID index is also constructed for fixed-length columns.
This enables efficient cross-column page alignment:
when a fused kernel accesses multiple columns of different lengths for the same set of RIDs, it can use binary search on each column's RID index to locate the corresponding pages efficiently.

\section{Implementation}
\label{sec:impl}

This section describes our prototype implementation of the DPF-based GPU-driven database engine (DPFProto).
We first describe the query execution engine that runs on GPU, then the data loader that prepares the on-storage layout.

\subsection{Query Engine}
\label{sec:impl-engine}

The engine reads compressed columnar pages from storage and executes IO, decompression, and relational operators within a single CUDA kernel, preceded by an upfront GPU-side pruning step.

\noindent
\textbf{BaM-based IO.}
We built on BaM~\cite{BAM23}, which exposes NVMe submission/completion queue pairs in GPU memory, allowing any GPU thread to submit storage requests without CPU involvement.
On top of BaM's primitives we implemented a thin layer that translates logical page identifiers into NVMe read commands:
given a page ID, the layer resolves the target SSD (pages are striped across SSDs in round-robin order) and the corresponding LBA range, constructs the command, and enqueues it via BaM's submission queue.

\noindent
\textbf{Integer decompression.}
For fixed-width integer columns, DPF requires a decompression routine callable as a device function that can place decoded values directly into shared memory.
We adopted GPU-FOR~\cite{DBLP:conf/sigmod/ShanbhagYYM22},
reusing its 32-bit routine for \texttt{INT32} and short \texttt{CHAR} columns ($n \le 2$), and extending it to 64-bit for \texttt{BIGINT}.
The 64-bit extension uses 64-bit references, modulo-64 bit indexing, and a 128-bit decode window; mini block boundaries remain word-aligned with at most $1/b$ padding overhead when the bit width $b$ is odd.

\noindent
\textbf{String decompression.}
For variable-length string columns, DPF adopts FSST~\cite{DBLP:journals/pvldb/Boncz0L20},
whose 255-entry static symbol table enables independent per-record decoding without cross-record state.
The 2,296-byte symbol table is small enough to reside in shared memory alongside the compressed payload and a staging buffer.
Each mini block is processed by one thread block (128 threads) using a two-pass structure:
the first pass computes per-record decompressed lengths and derives output offsets via an in-shared-memory prefix sum;
the second pass decodes records in parallel into the staging buffer, which is then written to global memory in warp-coalesced stores.
Existing GPU FSST implementations~\cite{DBLP:journals/sigops/VonkHA25,anema2025high} target bulk file decompression without per-record offsets within a split, providing neither the record-level decoding nor the shared-memory placement that DPF requires.

\noindent
\textbf{Fused kernel composition.}
The building blocks above are executed together within a single CUDA kernel.
Each thread block repeatedly picks up one logical page at a time,
driving the full IO--decompress--operate path end-to-end before advancing to the next page;
hence the per-block working set is bounded by a single page's columns rather than the table size.
For each page, the block batch-submits BaM IO requests for the required columns, polls their completions together, then invokes the per-column decompression routines in sequence (sharing a single shared-memory region via a union, since they run
sequentially).
Threads then iterate over column values in the page in parallel,
applying the operator chain -- filters, hash probes, and hash inserts -- directly on the decoded values, which are placed in shared memory, if they fit in it, or in GPU global memory.

\noindent
\textbf{Variable-length scan operators.}
The variable-length page layout (\S\ref{sec:design:compression}) enables scan operators over \texttt{varchar} attributes within the fused kernel.
Each thread decodes a record from its mini block and applies the operator to the decoded bytes without writing them back to global memory.
Our prototype implemented \texttt{LIKE}-style predicates by integrating a GPU Knuth--Morris--Pratt matcher~\cite{DBLP:journals/vldb/SitaridiR16} as a device function invoked on records emitted by the FSST routine.

\subsection{Data Loader}
\label{sec:impl-loader}

We implemented a data loader to prepare the on-storage layout aligned with DPF, in particular, type-specific compression.

\noindent
\textbf{Sort-based clustering and zone map building.}
Sort-based clustering with zone maps is a very typical organization scheme for analytical database~\cite{DBLP:conf/vldb/Moerkotte98,DBLP:journals/pvldb/OrrKC19}.
The prototyped data loader sorts all the incoming rows and places their column values in storage space in the key order,
while it can build a zone map for each column;
the map holds min/max values of the column and its associated columns (in the original row) for each page.
Our prototype implemented a capability of building reference-based zone maps.
the map is additionally allowed to hold min/max values of columns that are stored in the row to be joined with the original column.
This function was implemented in commercial database engines~\cite{DBLP:journals/pvldb/ZiauddinWKLPK17} and studied in recent literature on GPU-driven engines~\cite{GOLAP24}.

\noindent
\textbf{Multi-threaded two-pass loading.}
Our prototype writes each sorted table's columns to storage in two passes.
In the first pass, the loader determines the number of uncompressed pages per (column, file) pair and assigns each pair a contiguous region on the block device; regions of the same column are placed adjacently.
In the second pass, worker threads compress pages and write them densely from the start of each region, recording per-page zone maps (min/max values) for designated pruning attributes.

\noindent \textbf{Metadata.}
Each column carries three small side-files:
a page-offset array and a page-size array for $O(1)$ page location, and a prefix-sum array of per-page row counts for cross-column page alignment and post-pruning page-list intersection.
At a 1 MB page size and 1\,TB of data, these total roughly 20 MB (${\approx}\,2 \times 10^{-3}\%$ of the data volume).

Note that the data loader only runs on the host side because it was developed to organize the database aligned with DPF.
Efficient GPU-driven implementation remains a topic for future research.

\subsection{Current Limitations}
\label{sec:impl-limitations}

The current prototype has several limitations that point to natural directions for future work.
Joins are supported only when the build-side hash table fits in the GPU memory.
Storage access goes through BaM and is therefore restricted to raw block devices.
GPU-FOR and FSST are integrated as representative type-specific compression schemes; other schemes are not yet covered.
Among variable-length scan operators, only the \texttt{LIKE}-style predicate matcher is currently implemented.
\texttt{DECIMAL(18,\,2)} values are stored as 32-bit integers, but processed as 64-bit ones at query time; general fixed-point or floating-point decimal types are not supported.


\begin{figure*}[t!]
  \centering
  \begin{minipage}[b]{0.34\linewidth}
    \includegraphics[width=\linewidth]{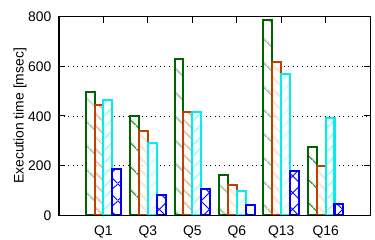}
    \subcaption{Query response (TPC-H).}
  \end{minipage}
  \begin{minipage}[b]{0.62\linewidth}
    \includegraphics[width=\linewidth]{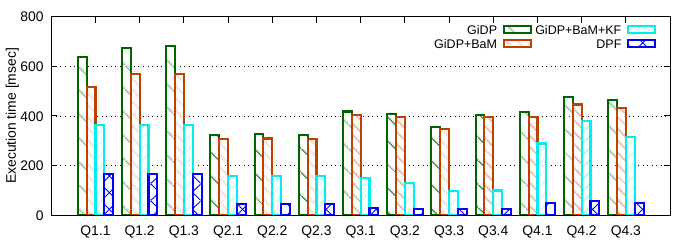}
    \subcaption{Query response (SSB).}
  \end{minipage}
  \\
  \begin{minipage}[b]{0.34\linewidth}
    \includegraphics[width=\linewidth]{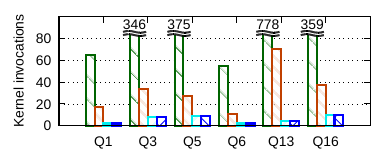}
    \subcaption{Kernel invocations (TPC-H).}
  \end{minipage}
  \begin{minipage}[b]{0.62\linewidth}
    \includegraphics[width=\linewidth]{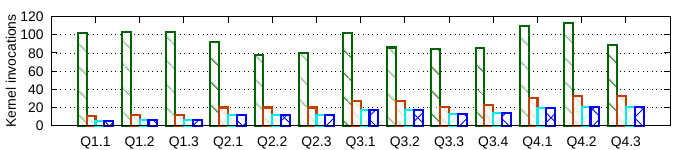}
    \subcaption{Kernel invocations (SSB).}
  \end{minipage}
  \\
  \begin{minipage}[b]{0.34\linewidth}
    \includegraphics[width=\linewidth]{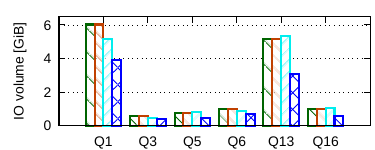}
    \subcaption{Total IO volume (TPC-H).}
  \end{minipage}
  \begin{minipage}[b]{0.62\linewidth}
    \includegraphics[width=\linewidth]{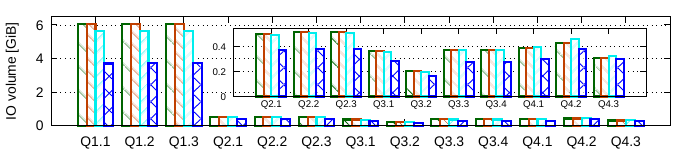}
    \subcaption{Total IO volume (SSB).}
  \end{minipage}
  \caption{Comparison of end-to-end query response, kernel invocations and total IO volume on TPC-H (left column) and SSB (right column).
  \textit{
  DPF achieved significant speedups over the baseline case:
  (a) 2.66 to 6.22 times speedups for six standard queries on TPC-H,
  and (b) 3.84 to 16.81 times speedups for thirteen standard queries on SSB.
  }
  }
  \label{res:queryresponse}
\end{figure*}

\section{Evaluation}
\label{sec:eval}

This section reports our experimental evaluation that we intensively and extensively conducted to demonstrate that \emph{Data Path Fusion} (DPF) performs better for analytical database workloads.
After showing our experimental setup, we present how the core technologies that make up DPF contribute to the performance improvements.
Then, we experimentally examine the impact of workload and configuration settings on DPF, as well as the overhead involved with data loading.
Finally, we compare DPF with other major implementations in terms of overall query performance.

\subsection{Experimental Setup}

We implemented a prototype of DPF-based GPU-driven database engine (DPFProto). 
For comparative studies,
the prototype supports the following test cases.
    \textbf{GiDP (baseline)} executes all the data path work on GPU;
    IO, pruning, decompression, and query operations are executed in separate GPU kernels invoked by the host.
    This case is treated as the baseline in this paper and corresponds to GOLAP, state-of-the-art work~\cite{GOLAP24}%
    \footnote{Since the original paper~\cite{GOLAP24} did not disclose the source code,
    we optimized the implementation as much as possible based on the disclosed information for comparison fairness.
    We mention two points regarding the implementation details.
    First, \textbf{GiDP (baseline)} implemented IO coalescing to merge multiple consecutive IOs and we conducted a page-size sensitivity analysis (presented in Section~5.4) since the original paper reported that larger page sizes offered higher performance.
    Second, our preliminary experiment confirmed that Deflate always performed poorer and thus we excluded Deflate from the code, even though the original paper~\cite{GOLAP24} reported the best of LZ4, Snappy and Deflate. 
    }.
    \textbf{GiDP+BaM} incorporates \emph{BaM} into the \textbf{GiDP (baseline)} case.
    BaM manages all IOs between GPU and storage instead of GDS.
    \textbf{GiDP+BaM+KF} further supports the \emph{kernel fusion} architecture;
    IO, pruning, decompression, and query operations are all executed in a single GPU kernel.
    Warp configurations in the kernel (e.g., the number of IO-dedicated warps versus decompression/operation warps) were tuned per query based on its IO-versus-compute balance.
    Due to the implementation limitation, the compression scheme only supports LZ4 and plain (non-compressed).
    \textbf{DPF} further incorporates the \emph{type-specific compression} capability.
    Obviously, this case corresponds to the full-fledged implementation of the DPF paradigm that this paper proposes.
This section highlights the performance difference among these cases to demonstrate how each building block of DPF offers speedups.

We utilized two standard benchmarks, the TPC-H benchmark (TPC-H)~\cite{tpch} and the star-schema benchmark (SSB)~\cite{DBLP:conf/tpctc/ONeilOCR09}.
These benchmarks have a strong benefit of being utilized in the state-of-the-art paper~\cite{GOLAP24} as well as in many other papers on database engines.
We extensively varied data scales and workload settings to verify their effects.
By default, the dataset was generated for TPC-H and SSB with a scale factor of 100, where the raw data sizes were approximately 107 GB and 62 GB, respectively.
A cold start was performed for each measurement, invalidating the cache throughout the IO path and the query execution path.
This paper reports an average value of ten trials.

Following the major known practice~\cite{GOLAP24,DBLP:conf/sigmod/SunFKX14,DBLP:conf/sigmod/YangCWGLMLKA20,DBLP:journals/pvldb/sunfww16},
we generated zone maps for all test cases and configurations in advance.
Zone maps were built only on the major tables --- LINEITEM and ORDERS for TPC-H (accounting for nearly 80\% of the total data), and LINEORDER for SSB (nearly 95\%);
the remaining dimension tables were small enough, not needing pruning.
For each of the major tables,
we built a zone map for each of the columns of the table and the columns of tables that can join with the table,
only if the column was evaluated with any comparative operators in predicates in test queries and its cardinality was sufficiently small (i.e., 5,000 or smaller in our test).

We conducted all the experiments on a modern machine,
having two Intel Xeon Gold 5418Y processors (24 cores running at 2.0 GHz in each), 512 GB host memory, and one NVIDIA A100 PCIe GPU (80 GB HBM2e device memory)
with one Samsung PM983 960 GB SSD and four Samsung PM9A3 1.92 TB SSDs.
The machine ran Ubuntu Linux 22.04, CUDA 12.9, and our prototype of DPF-based GPU-driven database engine (DPFProto).
The OS and software were stored in the file system volume organized in the PM983 space,
whereas the database was built on the database volume organized by four PM9A3 SSDs with the software striping capability (i.e., Linux mdadm).
The database page size, the work memory size in GPU, and the number of host-side threads were set to 1 MB, 40 GB%
\footnote{We followed the configuration disclosed in the paper~\cite{GOLAP24}}, and 32, respectively, unless otherwise noted.

\begin{figure*}[t!]
  \centering
  \begin{minipage}[b]{0.19\linewidth}
    \includegraphics[width=\linewidth]{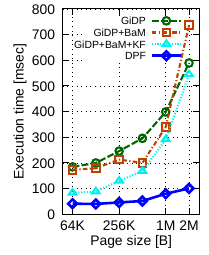}
    \subcaption{TPC-H Q3.}
  \end{minipage}
  \begin{minipage}[b]{0.19\linewidth}
    \includegraphics[width=\linewidth]{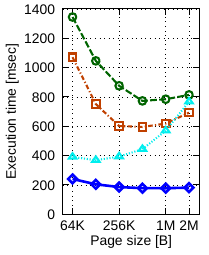}
    \subcaption{TPC-H Q13.}
  \end{minipage}
  \begin{minipage}[b]{0.19\linewidth}
    \includegraphics[width=\linewidth]{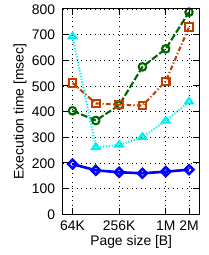}
    \subcaption{SSB. Q1.1}
  \end{minipage}
  \begin{minipage}[b]{0.19\linewidth}
    \includegraphics[width=\linewidth]{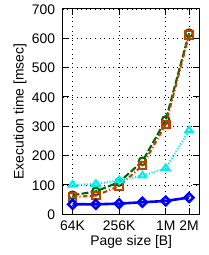}
    \subcaption{SSB. Q2.1}
  \end{minipage}
  \begin{minipage}[b]{0.19\linewidth}
    \includegraphics[width=\linewidth]{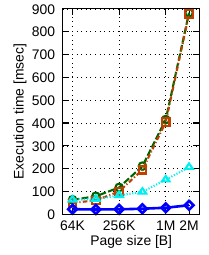}
    \subcaption{SSB. Q3.1}
  \end{minipage}
  \caption{Sensitivity analysis of page sizes
  \textit{
  \textbf{DPF} remains robust across all page sizes, offering speedups of 1.89 to 22.17 over the baseline case (\textbf{GiDP}).
  }
  }
  \label{res:pagesize}
\end{figure*}

\begin{figure*}[t!]
  \centering
  \begin{minipage}[b]{0.19\linewidth}
    \includegraphics[width=\linewidth]{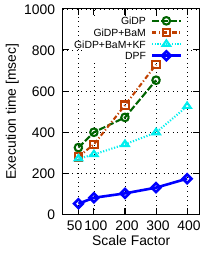}
    \subcaption{TPC-H Q3.}
  \end{minipage}
  \begin{minipage}[b]{0.19\linewidth}
    \includegraphics[width=\linewidth]{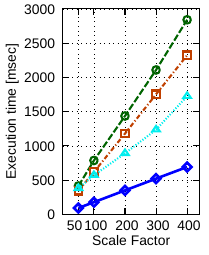}
    \subcaption{TPC-H Q13.}
  \end{minipage}
  \begin{minipage}[b]{0.19\linewidth}
    \includegraphics[width=\linewidth]{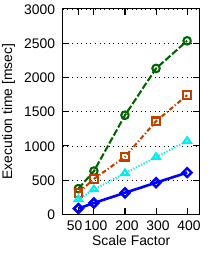}
    \subcaption{SSB Q1.1.}
  \end{minipage}
  \begin{minipage}[b]{0.19\linewidth}
    \includegraphics[width=\linewidth]{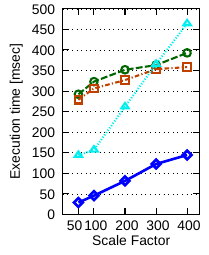}
    \subcaption{SSB Q2.1.}
  \end{minipage}
  \begin{minipage}[b]{0.19\linewidth}
    \includegraphics[width=\linewidth]{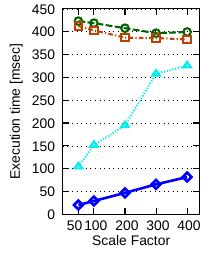}
    \subcaption{SSB 3.1.}
  \end{minipage}
  \caption{
  Data scalability analysis.
  \textit{
  \textbf{DPF} remains robust across all scale factors, offering speedups of 2.72 to 20.69 over the baseline case (\textbf{GiDP}).
  }
  }
  \label{res:datascalability}
\end{figure*}

\subsection{End-to-end Query Response}

We empirically studied the overall query speedups offered by DPF,
and investigated the impact of each of its core technologies.

Figure~\ref{res:queryresponse} shows that \textbf{DPF} achieved significant speedups over the baseline case (\textbf{GiDP}):
2.66 to 6.22 times speedups for six standard queries on TPC-H,
and 3.84 to 16.81 times speedups for thirteen standard queries on SSB.

Figure~\ref{res:queryresponse}~(a) summarizes the end-to-end query response of four different test cases for each of the six standard queries on TPC-H.
We chose these six queries because Q1, Q3, and Q5 were reported in the state-of-the-art paper~\cite{GOLAP24} and Q6, Q13 and Q16 were added by us since Q6 was the simplest scan query and Q13 and Q16 contained predicates on variable-length attributes.
\textbf{GiDP+BaM} shows that the use of BaM offered consistent query speedups of 1.12 to 1.51 in comparison with the baseline case (\textbf{GiDP}) using GDS.
\textbf{GiDP+BaM+KF} shows that the kernel fusion architecture showed speedup factors ranging from 0.95 to 1.23 over \textbf{GiDP+BaM} except for Q16;
Q16 was exceptional in that it exhibited a significant degradation.
\textbf{DPF} shows that the type-specific compression offered further speedups of 2.35 to 8.84 over \textbf{GiDP+BaM+KF} for all test queries including Q16,
where type-specific compression substantially reduced IO volume and faster decompression overcame the slowdown observed in \textbf{GiDP+BaM+KF}.
In sum, \textbf{DPF} successfully performed significantly faster (by factors of 2.66 to 6.22) than the baseline case (\textbf{GiDP}) across all queries.

\begin{figure*}[t!]
  \centering
  \begin{minipage}[b]{0.33\linewidth}
    \includegraphics[width=\linewidth]{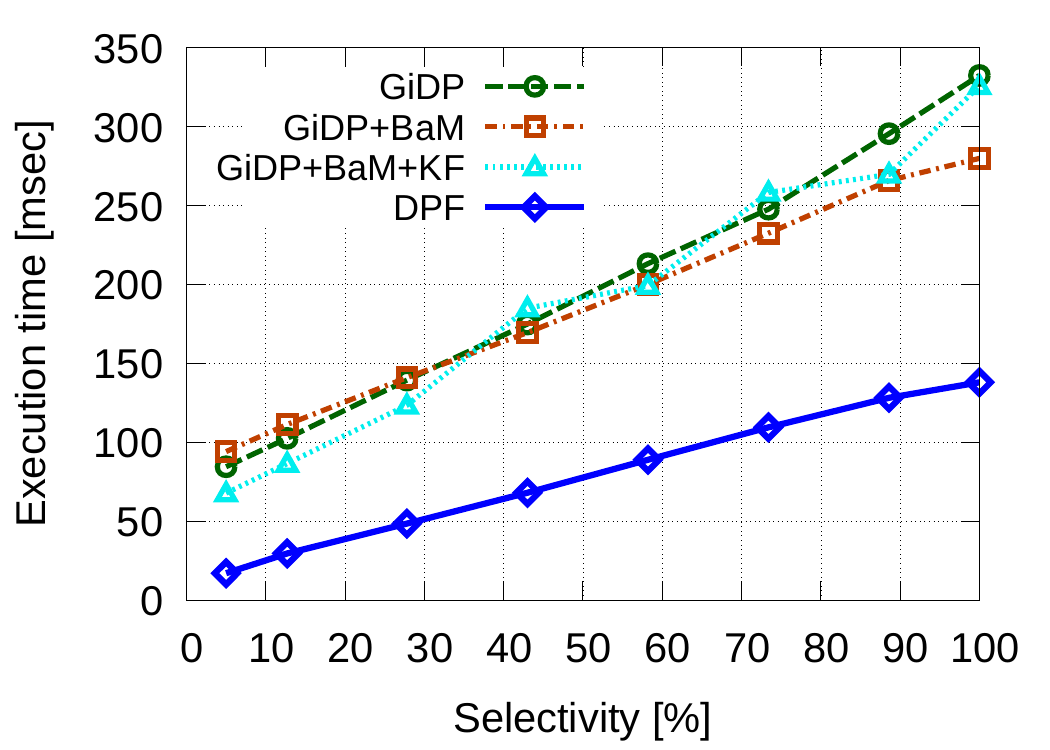}
    \subcaption{TPC-H Q6.}
  \end{minipage}
  \begin{minipage}[b]{0.33\linewidth}
    \includegraphics[width=\linewidth]{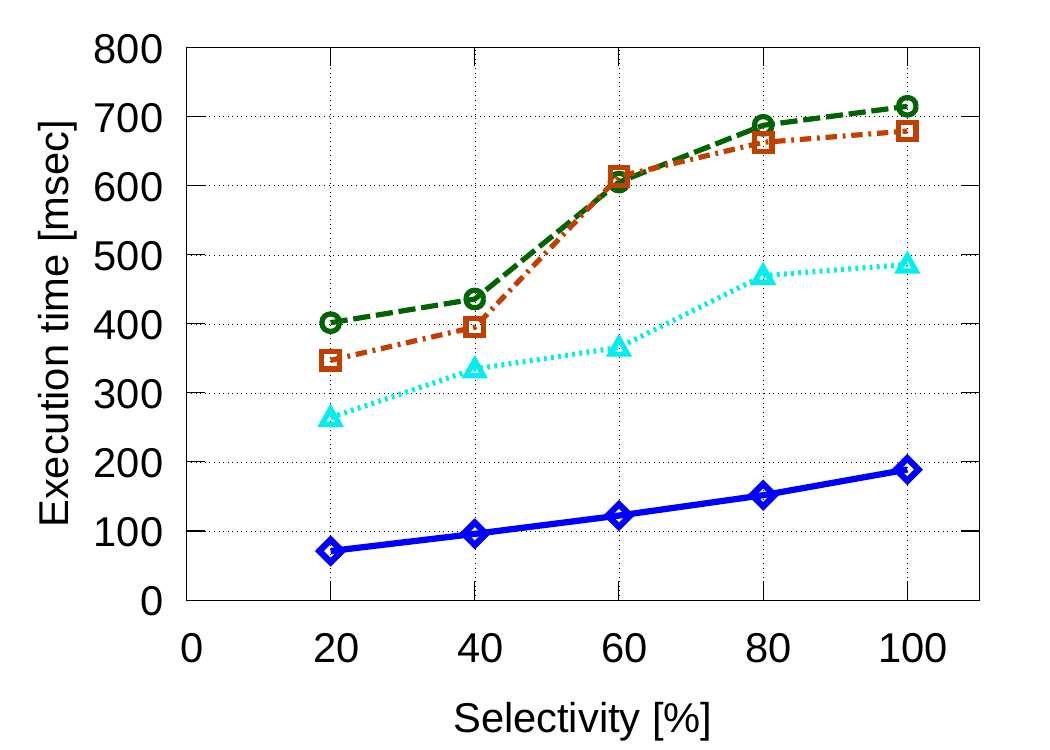}
    \subcaption{TPC-H Q3.}
  \end{minipage}
  \begin{minipage}[b]{0.33\linewidth}
    \includegraphics[width=\linewidth]{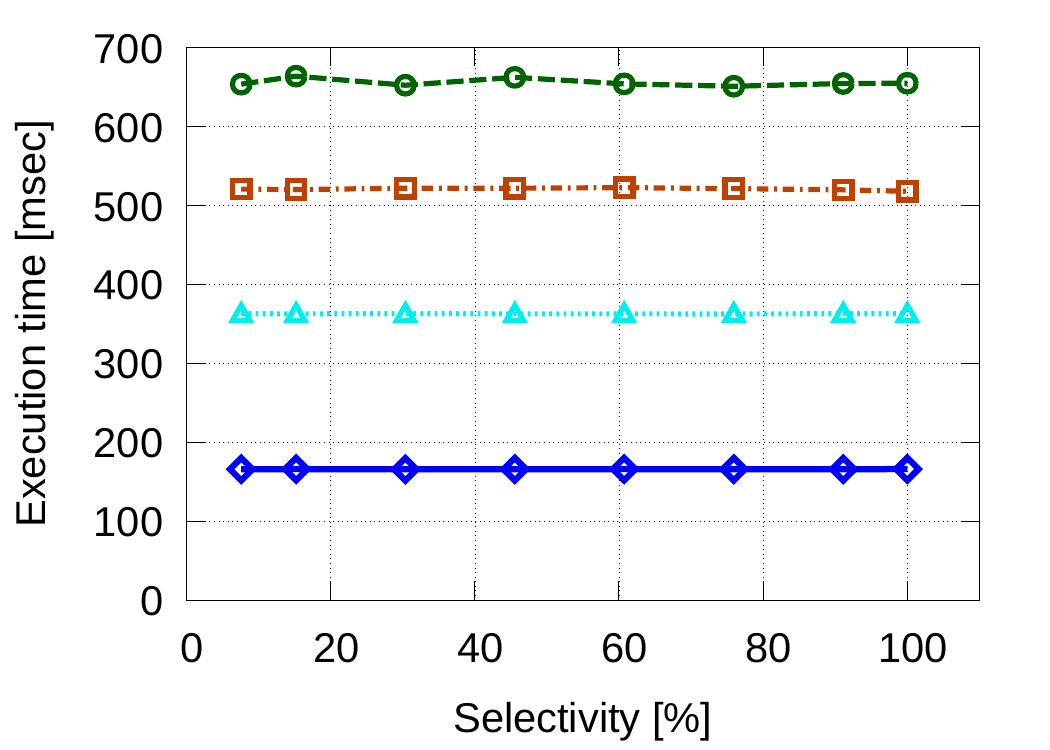}
    \subcaption{SSB Q1.1.}
  \end{minipage}
  \caption{
  Query selectivity sensitivity analysis.
  \textit{
  \textbf{DPF} provides consistent speedups of 2.26 to 5.66 across all selectivity values, regardless of whether pruning is effective for the query predicate.
  }
  }
  \label{res:queryselectivity}
\end{figure*}

Figure~\ref{res:queryresponse}~(c) reports the amount of kernel invocation during query execution.
\textbf{GiDP+BaM} shows that the use of BaM reduced the number of kernel invocations to 0.07 to 0.26 of that of \textbf{GiDP}; this is because a single host-side thread can initiate multiple IOs through a BaM kernel and multiple computation kernels on GPU, eliminating the need for separate kernel invocations for each IO.
Moreover, \textbf{GiDP+BaM+KF} and \textbf{DPF} show that the kernel fusion architecture and the type-specific compression consistently reduced the number of kernel invocations to 0.01 and 0.04, respectively; this is because a single fused kernel performs IO, decompression and query operators altogether.
Next, Figure~\ref{res:queryresponse}~(e) reports the total IO volume during query execution.
\textbf{GiDP+BaM} produced an IO volume mostly identical to that of \textbf{GiDP}, since both used the same compression scheme selected by the loader.
\textbf{GiDP+BaM+KF} produced moderately different IO volumes (0.79 to 1.12 of \textbf{GiDP})%
\footnote{This was caused by the implementation limitation; \textbf{GiDP+BaM+KF} only supported LZ4 and plain as compression schemes.}.
\textbf{DPF} shows that the type-specific compression successfully reduced the IO volume to 0.55 to 0.71 of that of GiDP.

Figure~\ref{res:queryresponse}~(b), (d) and (f) summarize the same experiments for the thirteen standard queries on SSB.
The observation was similar to the cases on TPC-H.
Figure~\ref{res:queryresponse}~(b) presents the impact of each core technology to the overall query speedup.
The use of BaM offered consistent query speedups of 1.02 to 1.23 in comparison with the baseline case,
the kernel fusion architecture offered further query speedups of 1.17 to 3.97, and then
the type-specific compression offered consistent query speedups of 2.19 to 6.63.
Taken together, \textbf{DPF} successfully performed 3.84 to 16.81 times faster than the baseline case.
Figure~\ref{res:queryresponse}~(d) and (f) present the reduction on kernel invocations and IO volume.
\textbf{DPF} successfully reduced kernel invocations to 0.05 to 0.24 in comparison with the baseline,
and reduced IO volume by factors of 0.61 to 0.97.

\subsection{Page-size Sensitivity}

The page is a critical unit for IO and processing parallelization in GPU-driven database engines; it might affect the overall query processing performance.
We analyzed page-size sensitivity
by conducting the same performance test as presented in the previous subsection
while changing page sizes from 64 KB to 2 MB.
This section highlights only Q3 and Q13 for TPC-H and Q1.1, Q2.1 and Q3.1 for SSB;
we observed similar results with other queries.

Figure~\ref{res:pagesize} summarizes the end-to-end response of five different test queries, indicating that \textbf{DPF} remains robust across all page sizes, offering speedups of 1.89 to 22.17 over \textbf{GiDP}, the baseline case.
Although the degree varied depending on the query, in \textbf{GiDP} and its variants,
the response time exhibited a concave-down curve.
This suggests that query processing performance was strongly influenced by the page size setting.
To make matters worse, the lowest point of the curve varied depending on the query.
In other words, the optimal page size differs depending on the query.
On the other hand, a similar trend was observed for \textbf{DPF}, though to a much lesser extent.
The impact of the page size setting on query response time was significantly mitigated.
This was because type-specific compression/decompression allows fine-grained parallel processing of columns, more intensively feeding GPU's computational capacity.

\subsection{Data Scalability}

Scalability with respect to data size is a key requirement for database engines, especially when the database does not fit in memory and must be stored in secondary storage.
This paper presents our experimental study on dataset scalability,
in which we conducted the same performance test while changing dataset scales.

Figure~\ref{res:datascalability} summarizes the end-to-end query response.
It indicates that \textbf{DPF} remains robust across all scale factors, offering speedups of 2.72 to 20.69 over \textbf{GiDP}.
Overall, the query took longer for larger data scales, as was often seen.
Notably, \textbf{DPF} always performed significantly better than the other cases,
offering 1.76 to 2.72 times speedups \emph{even} over the other best case.

\begin{table}[t]
    \centering
    \caption{Data loading time comparison.
    \textit{
    \textbf{DPF} needs merely comparable loading time with the baseline case (\textbf{GiDP}).
    }
    }
    \label{tab:loading}
    \small
    \begin{tabular}{l|rrrr|rrrr}
        \toprule
        & \multicolumn{4}{c|}{TPC-H}
        & \multicolumn{4}{c}{SSB} \\
        \cmidrule(lr){2-5}\cmidrule(lr){6-9}
        [s]
        & 50 & 100 & 200 & 300
        & 50 & 100 & 200 & 300 \\
        \midrule
        GiDP & 1556 & 3176 & 9420 & 16050 & 3312 & 6623 & 21326 & 38724 \\
        DPF  & 1544 & 3177 & 9429 & 16067 & 3262 & 6538 & 21173 & 38495 \\
        \bottomrule
    \end{tabular}
\end{table}

\begin{table*}[t!]
    \centering
    \caption{Performance comparison among our prototype of DPF-based GPU-driven database engine and other major database engines.
    \textit{
    DPF consistently achieves significant speedups:
    11.2 to 176.8 times faster for TPC-H and 7.3 to 123.3 times faster for SSB than Polars and Spark-RAPIDS;
    9.3 to 81.8 times faster for TPC-H and 8.8 to 102.5 times faster for SSB than DuckDB.
    }
    }
    \label{tab:majorimpl}
    \footnotesize
    \begin{tabular}{l|rrrrrr|rrrrrrrrrrrrr}
        \toprule
        & \multicolumn{6}{c|}{TPC-H (SF=100)}
        & \multicolumn{13}{c}{SSB (SF=100)} \\
        \cmidrule(lr){2-7}\cmidrule(lr){8-20}
        [ms]
        & Q1  & Q3  & Q5   & Q6  & Q13 & Q16
        & Q1.1 & Q1.2 & Q1.3 & Q2.1 & Q2.2 & Q2.3
        & Q3.1 & Q3.2 & Q3.3 & Q3.4 & Q4.1 & Q4.2 & Q4.3 \\
        \midrule
        \textbf{DPF (GPU)}
        & \textbf{187}  & \textbf{81}  & \textbf{104}   & \textbf{42}   & \textbf{178}  & \textbf{44}
        & \textbf{166}  & \textbf{166}  & \textbf{166}  & \textbf{46}   & \textbf{46}   & \textbf{46}
        & \textbf{29}   & \textbf{24}   & \textbf{26}   & \textbf{26}   & \textbf{51}   & \textbf{58}   & \textbf{48}  \\      Polars (GPU)
        & 3371 & 2520 & 3543 & 1563 & 2178 & 498
        & 1524 & 1511 & 1508 & 2766 & 2730 & 2743
        & 3082 & 3003 & 2979 & 2978 & 3576 & 3364 & 3363 \\
        Spark-RAPIDS (GPU)
        & 3882 & 14266 & 11993 & 885 & 3037 & 2777
        & 1264 & 1235 & 1215 & 3134 & 2118 & 1989
        & 2188 & 2100 & 1996 & 1883 & 3350 & 3005 & 2925 \\
        \midrule
        DuckDB (CPU)
        & 1732 & 6603 & 3056 & 1004 & 2240 & 814
        & 1475 & 1466 & 1460 & 2280 & 2238 & 2278
        & 2557 & 2498 & 2488 & 2470 & 2987 & 2946 & 2896 \\
        \bottomrule
    \end{tabular}
\end{table*}

\subsection{Query Selectivity}

Query performance is affected by the selectivity on tables to be scanned on storage,
in particular, when pruning techniques are employed.
We analyzed query selectivity
by conducting the similar performance test
while changing selection conditions of each test query.
Specifically, we tested the query performance while varying the predicate condition on \texttt{L\_SHIPDATE} for Q6, \texttt{C\_MKTSEGMENT} for Q3 on TPC-H, and \texttt{LO\_ORDERDATE} for Q1.1 on SSB.

Figure~\ref{res:queryselectivity} summarizes the end-to-end query response,
showing that \textbf{DPF} provides consistent speedups of 2.26 to 5.66 across all selectivity values, regardless of whether IO pruning is effective to the query predicate.
We observed that the query response time increased as the selectivity ratio increased for Q6 and Q3 on TPC-H in all cases from \textbf{GiDP} to \textbf{DPF}.
This was likely because the zone map was aligned for the query predicate and thus the pruning effectively worked for the query.
For these workloads, \textbf{DPF} consistently performed much faster than the others (2.26 to 5.66 times faster than \textbf{GiDP}).
In contrast, looking at Q1.1 on SSB,
the query response time did not change as the selectivity ratio changed.
This was probably caused by the zone map not being aligned with the query predicate; the pruning was not effective for the query.
Even in this case, \textbf{DPF} consistently performed much faster than the others (3.93 to 4.00 times faster than \textbf{GiDP}).

\subsection{Loading Cost}

DPF modifies physical database design on secondary storage to incorporate type-specific compression, likely increasing the layout complexity.
It is essential to verify how significant this modification's impact is on the time overhead required for database loading.
We present the experimental study on the loading time.
We generated datasets with different scale factors for TPC-H and for SSB, respectively,
and then tested the database loading performance.

Table~\ref{tab:loading} summarizes the loading time in different settings, indicating that \textbf{DPF} needs merely comparable loading time with \textbf{GiDP} across all the tested scale factors (50 to 300) of TPC-H and SSB.
Specifically, the additional loading cost for incorporating type-specific compression is negligible (i.e., within 1.5\%).

\subsection{Comparison with Other Implementations}

Finally, we present the comparative performance study between our prototype of the DPF-based GPU-driven database engine and other major database engines.
First, we tested two major GPU-driven database engines, Polars GPU Engine (Polars)~\cite{polars} and RAPIDS Accelerator for Apache Spark (Spark-RAPIDS)~\cite{spark-rapids}, 
since they ran query processing on GPU and supported GPUDirect Storage (GDS).
We also attempted to test other GPU-driven database engines, PG-Strom~\cite{pgstrom} and HeavyDB~\cite{heavydb}, but abandoned this idea
because PG-Strom required a commercial license to work with GDS for striped NVMe SSDs and HeavyDB was no longer maintained%
\footnote{We followed the community's convention of not disclosing performance metrics measured in commercial implementations.
HeavyDB appears to have been discontinued after its acquisition by NVIDIA in 2025.}.
Polars 1.35.2 with RAPIDS cuDF 26.02 and Spark-RAPIDS based on Apache Spark 4.1.1 with RAPIDS Accelerator 26.02 were configured to use 48 host threads and 72 GB of GPU memory.
Second, just for reference, we also tested a major CPU-based analytical database engine, DuckDB~\cite{DBLP:conf/sigmod/RaasveldtM19}, which ran query processing on CPU but had the benefit of having been experimentally tested in many recent papers.
DuckDB 1.4.4 was configured to use 48 host threads and 384 GB of host memory.
Note that Polars, Spark-RAPIDS and DuckDB were tuned honestly to ensure that they obtain their best achievable performance;
they actually utilized the same amount of resources (i.e., host threads and GPU memory) as our prototype or even more.

For each of these database engines,
we built the database in the same database volume used in the experiments described earlier.
Since these database engines could not mount block storage directly, we configured an XFS file system and built the databases on top of it.
All the tests were conducted on the same machine as presented so far.
We also tuned each database engine with known practices.

Table~\ref{tab:majorimpl} summarizes the comparison of the end-to-end query response of DPF and other database engines.
Obviously, DPF consistently achieved significant speedups over others.
Specifically, DPF performed 11.2 to 176.8 times faster for TPC-H and 7.3 to 123.3 times faster for SSB than Polars and Spark-RAPIDS, two major GPU-driven database engines.
This experimental result demonstrates that DPF efficiently utilizes the computational capacity and the IO bandwidth that the same GPU holds in comparison with other major implementations.
In addition, DPF performed 9.3 to 81.8 times faster for TPC-H and 8.8 to 102.5 times faster for SSB than DuckDB, a major CPU-based analytical database engine.
Since DPF and DuckDB ran on different processor architectures, it is not appropriate to directly compare their performance.
However, this observation shows that DPF delivers performance on par with or exceeding that of DuckDB,
suggesting that DPF has the potential to demonstrate superiority as a database engine.

\subsection{Summary}

Our experiments demonstrate that \textbf{DPF} consistently and significantly outperforms the state-of-the-art approach across all tested dimensions.
DPF achieved speedups of 2.66 to 6.22 times on TPC-H and 3.84 to 16.81 times on SSB over \textbf{GiDP (baseline)} in terms of end-to-end query response,
indicating that the combination of kernel fusion, BaM-based IO and type-specific compression contributes cumulatively to the improvement.
DPF remained robust under varying page sizes, data scales and query selectivities.
The loading cost required to organize the DPF storage layout was
negligible.
Finally, in comparison with other major database engines, \textbf{DPF} performed 7.3 to 176.8 times faster than GPU-driven engines,
confirming that \emph{Data Path Fusion} effectively unlocks the computational and IO potential of modern GPU for analytical query workloads.


\section{Related Work}
\label{sec:related}

GPU-accelerated data processing has been an active area of research for over two decades.
Several surveys~\cite{DBLP:journals/tlsdkcs/BressHSBS14,DBLP:journals/csur/RosenfeldBM23} and experimental characterizations~\cite{DBLP:journals/pvldb/CaoSIAK23} provide comprehensive overviews of the field.
We organize the closely related work into five categories below, and position DPF with respect to each.

\noindent
\textbf{GPU-based analytical query processing.}
Research on GPU-accelerated relational operators has progressed from individual operator studies to complete analytical query engines.
Join algorithms have received extensive attention~\cite{DBLP:conf/sigmod/HeYFLGLS08,DBLP:conf/icde/SioulasCKAA19,DBLP:journals/tpds/YabutaNKEK17,DBLP:conf/ssdbm/RuiT17,DBLP:journals/concurrency/BellasG17}, spanning multi-way and nested-loop variants~\cite{DBLP:journals/vldb/LaiSLX22,DBLP:conf/ieeehpcs/NguyenEK18}, join optimization for complex queries~\cite{DBLP:conf/icde/MageirakosMKCA22}, and revisits as GPU hardware evolved~\cite{DBLP:conf/damon/KaldeweyLMV12,DBLP:conf/bigdataconf/RuiLT15}.
Sorting~\cite{DBLP:conf/sigmod/GovindarajuGKM06,DBLP:conf/sigmod/SatishKCNLKD10,DBLP:conf/ipps/GressZ06,DBLP:conf/ipps/LeischnerOS10,DBLP:conf/sigmod/StehleJ17}, aggregation and query compilation~\cite{DBLP:journals/pacmmod/WuKA25,DBLP:conf/sigmod/FunkeBNMT18,DBLP:journals/pvldb/FunkeT20,DBLP:journals/vldb/BressKFZRM18,DBLP:conf/micro/WuDCY12}, index structures~\cite{DBLP:conf/sigmod/KimCSSNKLBD10,DBLP:conf/damon/BeierKS12}, nested query processing~\cite{DBLP:conf/icde/FloratosXWGYLZ21}, and string processing~\cite{DBLP:journals/vldb/SitaridiR16} have similarly been studied in depth.
Lightweight compression for GPU-resident data has been explored in several systems~\cite{DBLP:journals/pvldb/FangHL10,DBLP:conf/sigmod/ShanbhagYYM22,DBLP:journals/pvldb/HuangSLCCIDS25}, and Afroozeh et al. demonstrated the effectiveness of the FastLanes compression layout~\cite{DBLP:journals/pvldb/AfroozehB23} on GPUs~\cite{DBLP:conf/damon/AfroozehFB24,DBLP:conf/damon/HepkemaAFBM25}.
Another line of work repurposes hardware abstractions originally designed for other workloads:
TCUDB~\cite{DBLP:conf/sigmod/HuLT22} runs SQL operators on tensor cores, tensor computation runtimes have been used as query execution backends~\cite{DBLP:journals/pvldb/HeNBSPCCKI22}, Garuda~\cite{DBLP:journals/pvldb/DoraiswamyK0H23} maps relational operators to the graphics pipeline, and ray-tracing and rasterization primitives have been applied to database indexing and scan~\cite{DBLP:journals/pvldb/Henneberg023,DBLP:conf/icde/HennebergSKB25,DBLP:journals/pvldb/ShiZWZL25,DBLP:conf/cidr/DoraiswamyH26}.
At the engine level,
GPL~\cite{DBLP:conf/sigmod/PaulHH16} introduces efficient pipelined query execution, Themis~\cite{DBLP:journals/pvldb/HongKLMBH24} addresses warp-level load imbalance under skewed data, Pyper~\cite{DBLP:journals/pvldb/PaulHHL20} improves JIT-compilation efficiency, MultiQx-GPU~\cite{DBLP:journals/pvldb/WangZYMLD014} enables concurrent query execution on a shared GPU, and Crystal~\cite{DBLP:conf/sigmod/ShanbhagMY20}, Ocelot~\cite{DBLP:journals/pvldb/HeimelSPMM13}, Voodoo~\cite{DBLP:journals/pvldb/PirkMZM16}, and Sirius~\cite{DBLP:conf/cidr/YogatamaYKSKCTP26} provide complete engines and performance characterizations.
Several of these systems employ operator fusion to merge multiple relational operators into a single GPU kernel~\cite{DBLP:journals/pvldb/LiTLPS16,DBLP:conf/sigmod/ShanbhagMY20,DBLP:journals/pvldb/HongKLMBH24},
reducing intermediate materialization between operators.
DPF extends this philosophy further:
our kernel fusion architecture unifies not only database operators but also IO and decompression into a single kernel,
eliminating the need for separate kernel invocations across the entire data path.
DPF complements these efforts by shifting the focus from in-memory operator efficiency to end-to-end data path optimization that spans storage IO through query result production.

\noindent
\textbf{CPU and GPU co-execution.}
A substantial body of work has explored hybrid CPU-GPU query execution, including hash-join co-processing on coupled CPU-GPU architectures~\cite{DBLP:journals/pvldb/HeLH13,DBLP:journals/pvldb/HeZH14}, data placement and caching policies for heterogeneous systems~\cite{DBLP:journals/pvldb/YogatamaGY22,DBLP:journals/pvldb/YogatamaGY24}, parallel execution frameworks with JIT-compiled engines~\cite{DBLP:journals/pvldb/ChrysogelosKAA19}, robust query processing under scarce co-processor resources~\cite{DBLP:conf/sigmod/BressFT16}, adaptive work placement across heterogeneous computing resources~\cite{DBLP:journals/pvldb/KarnagelHL17}, index structures shared across CPU and GPU memory~\cite{DBLP:conf/sigmod/ShahvaraniJ16}, and hybrid strategies that filter on CPU to handle databases exceeding GPU memory~\cite{DBLP:journals/pvldb/LiDWMABBCIPRCG25}.
These approaches fundamentally rely on CPU as an active participant.
DPF takes a different stance by delegating the entire data path -- from IO initiation to query result production -- to GPU, minimizing host-side involvement.

\noindent
\textbf{Multi-GPU execution.}
Multi-GPU query processing over fast interconnects such as NVLink and RDMA has been explored for joins~\cite{DBLP:conf/sigmod/LutzBZRM20,DBLP:conf/sigmod/LutzBZRM22,DBLP:conf/sigmod/PaulLHL21,DBLP:journals/pvldb/RuiLT20,DBLP:journals/pacmmod/ThostrupDBLB23,DBLP:journals/pvldb/MaltenbergerITR25} and sorts~\cite{DBLP:conf/sigmod/MaltenbergerITR22}.
Vortex~\cite{DBLP:journals/pvldb/YuanIMT24} aggregates PCIe links across multiple GPUs to feed a single target GPU for workloads exceeding GPU memory.
Recent work also examines scale-out analytics at terabyte scale~\cite{DBLP:journals/pvldb/WuCCIS25} and benchmarks GPU and interconnect combinations across relational workloads~\cite{DBLP:journals/pvldb/KabicWDA25}.
DPF currently targets single-GPU execution; extending it to multi-GPU configurations is an interesting direction for future work.

\noindent
\textbf{Efficient IO operations for GPU.}
When the working set exceeds GPU memory and data resides on storage, IO efficiency becomes a critical bottleneck.
HippogriffDB~\cite{DBLP:journals/pvldb/LiTLPS16} demonstrated direct NVMe-to-GPU transfers via peer-to-peer DMA with GPU-side decompression.
GOLAP~\cite{GOLAP24} advanced this direction with a GPU-in-Data-Path architecture combining GPUDirect Storage-based IO with GPU-side pruning and on-the-fly decompression.
HetCache~\cite{DBLP:conf/cidr/NicholsonRCA23} synergizes NVMe storage with GPU acceleration for memory-efficient analytics, and Nicholson et al.~\cite{DBLP:conf/damon/NicholsonCBA25} analyzed how compression affects GPU-accelerated query execution on out-of-memory datasets.
These systems rely on CPU-initiated IO (e.g., GPUDirect Storage), where the host orchestrates data transfers.

In contrast, GPU-initiated storage access provides an alternative
model.
BaM~\cite{BAM23} introduced a foundational primitive for this approach, enabling GPU threads to submit storage requests directly through software cache and high-throughput queues.
Torp et al.~\cite{DBLP:conf/damon/TorpLT25} benchmarked BaM against CPU-centric storage interfaces, and AGIO~\cite{DBLP:conf/asplos/HanSCMQH26} introduced asynchronous IO APIs that decouple IO initiation from completion to hide SSD access
latency.
NVIDIA has presented SCADA~\cite{nvidia-scada}, extending this line with a programming model for high-throughput, fine-grained, GPU-initiated access.
Related applications of GPU-driven IO include GNN sampling and training~\cite{DBLP:journals/pvldb/ParkMQH24,Hyperion25}, large-scale similarity
search~\cite{FlashANNS/10.1145/3786652}, and batched storage access coordination~\cite{DBLP:conf/icde/SongZSSYZCWTW25}.
Complementary work on GPU-aware file systems exposes file-system abstractions to GPU kernels~\cite{DBLP:journals/tocs/SilbersteinFKW14,DBLP:journals/sigops/ShaharBS18,DBLP:conf/fast/QiuLHYSYCZZ25}, and DRAGON~\cite{DBLP:conf/sc/MarkthubBLVM18} extends CUDA unified memory to NVM via the GPU page-faulting mechanism.
DPF builds on the BaM primitive and integrates GPU-initiated IO directly into the fused kernel, allowing IO, decompression, and query operations to execute without yielding control to the host.

\noindent
\textbf{Other database workloads on GPU.}
Beyond analytics, GPUs have been applied to OLTP transaction processing~\cite{DBLP:conf/sigmod/BoeschenB22,DBLP:conf/osdi/QianG24,DBLP:journals/pvldb/HeY11,DBLP:conf/icde/Wei0L0LZJ024}, graph analytics including traversal, subgraph enumeration, and subgraph matching~\cite{DBLP:conf/sigmod/ShaLT19,DBLP:conf/sigmod/GuoL0HXT20,DBLP:journals/pacmmod/Sun023}, matrix computation engines~\cite{DBLP:conf/sigmod/HanNLPKK19,DBLP:conf/sigmod/HanL022}, spatial and moving-object queries~\cite{DBLP:journals/pvldb/WangHLWZS12,DBLP:journals/vldb/WardHZQ14}, and similarity search and skyline computation~\cite{DBLP:journals/pvldb/BoghCA15,DBLP:journals/vldb/TeodoroVMTMS14}.
These workloads present distinct execution patterns from analytical query processing; exploring how the data path fusion concept might extend to such domains is left for future investigation.


\section{Conclusion}
\label{sec:conclusion}

This paper proposed \emph{Data Path Fusion} (DPF), a GPU-driven data processing paradigm that fuses IO, decompression, and database operations into a single GPU kernel for analytical query processing.
By eliminating inter-kernel synchronization and intermediate materialization across the data path,
DPF allows GPU to sustain its massively parallel throughput from the moment data leaves storage to the moment query results are produced.
DPF incorporates type-specific compression that tailors encoding to each column's data type, variable-length attribute support that broadens applicability to real-world analytical schemata, and BaM-based IO that enables fine-grained, GPU-initiated storage access inherently compatible with the fused kernel architecture.
Our extensive experimental evaluation using TPC-H and SSB benchmarks demonstrated that DPF achieves significant speedups (2.66 to 6.22 times on TPC-H and 3.84 to 16.81 times on SSB in the representative configuration) over the state-of-the-art approach with performance robustness across a wide range of page sizes, data scales, and query selectivities plus negligible additional loading cost.

Several directions remain for future work.
Extending DPF to support out-of-core joins when hash tables exceed GPU memory, incorporating additional compression schemes beyond GPU-FOR and FSST, and scaling the fused kernel architecture to multi-GPU configurations are natural next steps toward a comprehensive GPU-native analytical database system.


\begin{acks}
This work has been in part supported by JST Support for Pioneering Research Initiated by the Next Generation (SPRING) JPMJSP2108, and Cross-ministerial Strategic Innovation Promotion Program (SIP) on Integrated Health Care System JPJ012425.
The authors used generative AI tools (Claude by Anthropic)
to assist in portions of the implementation, specifically for generating boilerplate and utility code during the experimental evaluation,
as well as to assist in surveying related literature. 
All AI-assisted outputs were carefully reviewed, verified, and revised by the authors, who take full responsibility for the content of this paper.
\end{acks}

\bibliographystyle{ACM-Reference-Format}
\bibliography{reference}

\end{document}